УДК 378.091:004.4

**Величко Владислав Євгенович**
канд. фіз-мат наук, доцент
ДВНЗ «Донбаський державний педагогічний університет», м. Слов'янськ, Україна
ORCID ID 0000-0001-9752-0907
*vladislav.velichko@gmail.com*

# СТВОРЕННЯ ЕЛЕКТРОННИХ НАВЧАЛЬНИХ КУРСІВ ЗАСОБАМИ ВІЛЬНОГО ПРОГРАМНОГО ЗАБЕЗПЕЧЕННЯ

**Анотація.** Інформатизація освіти збагачує традиційні методики навчання новими формами і методами, що ґрунтуються на широкому і гармонійному застосуванні ІКТ. Електронна педагогіка набуває статусу одного з найзатребуваніших напрямів розвитку педагогіки сьогодення, у якому такий засіб, як електронні навчальні курси, є акумулюванням сучасних інформаційно-комунікаційних технологій і теорії та практики електронного навчання. У статті розглянуто стандарти обміну інформацією між системами навчання і вільне програмне забезпечення, що використовується для створення електронних навчальних курсів. Наголошено, що електронні навчальні курси є як частиною дистанційних систем навчання, так і самостійним навчальним засобом. Розглянуті системи створення електронних навчальних курсів дають можливість створювати електронні освітні ресурси, що ґрунтуються на сучасному уявлені про об'єктну модель представлення інформації.

**Ключові слова:** електронне навчання; вільне програмне забезпечення; електронні навчальні курси; XerteOnlineToolkits; eLearning XHTML editor; ReloadEditor.

## 1. ВСТУП

Розширення форм організації, методів і засобів навчання, що ґрунтуються на використанні ІКТ, спонукає дослідників не тільки до їх утілення й апробацій, а й до системного аналізу їх можливостей, очікуваних переваг і можливих недоліків. На сьогодні науковцями і дослідниками зроблено багато спроб поєднання різноманітних педагогічних технологій, що ґрунтуються на використанні ІКТ. Отже, відбуваються формування й теоретична систематизація об'єктивних знань про дійсність; описуються, пояснюються і прогнозуються нові педагогічні явища, що виникають у процесі використання ІКТ в освітній діяльності, яка створює підґрунтя нового напряму в педагогіці – електронній педагогіці. У її межах відбувається створення та розробка системи електронного навчання, у якій серед існуючих засобів навчання використовуються електронні навчальні курси. Для створення електронних навчальних курсів застосовують як прикладне програмне забезпечення загального призначення, так і системи, створені спеціально. Спеціалізоване програмне забезпечення розробки електронних навчальних курсів – це складні інформаційні системи, що потребують капіталовкладень у їх розробку. Останнє відбувається двома шляхами. Перший полягає у використанні фінансових витрат на створення програмного продукту, а другий – у використанні добровільних інтелектуальних «внесків» у вигляді розробки вихідних кодів програмних продуктів від ентузіастів, які розробляють вільне програмне забезпечення.

**Постановка проблеми.** На сьогодні інновації у сфері освіти пов'язані насамперед з інформатизацією освітньої діяльності. Аналіз наукових джерел, результатів конференцій і семінарів дав можливість зробити висновок, що більшість досліджень ґрунтуються на поточній практиці використання інформаційно-комунікаційних технологій в освітній діяльності, накопиченні емпіричного досвіду та спробах створення науково-обґрунтованого базису через розробку й апробацію електронної педагогіки [1]. Одним із завдань електронної педагогіки є критичне осмислення нових видів навчальних занять, що стали результатом застосування інформаційно-комунікаційних технологій і створення нових підходів – таких, як конективізм, за якого навчання розглядається як процес створення мережі, вузлами якої є різноманітні сутності (люди, організації, бібліотеки, сайти, книги, журнали, бази даних і будь-які інші джерела інформації) [6]. Конективізм набув своєї значущості в освіті завдяки постійно зростаючому рівню розвитку ІКТ, що відповідає сутності нової методики проектування: системи стають кращими (повнішими, інформативнішими) за умови, якщо ними користуються якомога більше людей [9]. Залучення людей до наповнення контентом і багаторазова перевірка опублікованих матеріалів є відтворенням ідеї розробки вільного програмного забезпечення. Оскільки сучасні тенденції в навчанні еквівалентні підходам до розробки вільного програмного забезпечення, постає загальне питання про взаємодію вільного програмного забезпечення і сучасних підходів до проблеми електронного навчання. До такої взаємодії можна віднести практичне питання про використання вільного програмного забезпечення з метою створення засобів електронного навчання. Означена проблема була експериментально досліджена в процесі професійної підготовки майбутніх учителів математики на фізико-математичному факультеті ДВНЗ «Донбаський державний педагогічний університет».

**Аналіз останніх досліджень і публікацій.** В арсеналі електронного навчання наявні насамперед електронні освітні ресурси (ЕОР) [4]. Електронний навчальний курс відповідно до класифікації електронних освітніх ресурсів за своєю функціональною ознакою належить до навчальних ЕОР, за формою подання – до інформаційної системи, а за створенням має безпосереднє відношення до програмного продукту й належить до категорії «курс дистанційного навчання». Поза дистанційною формою навчання електронний навчальний курс може використовуватися також і в інших організаційних формах освіти, мати дидактичну значущість і створюватись окремо від систем організації дистанційної освіти. Дослідженню проблем розробки, упровадження, класифікації, систематизації електронних навчальних курсів присвячено роботи О. Глазунової, В. Кухаренка, Н. Морзе, Ю. Триуса та інших.

Електронні навчальні курси складаються зі значно менших об'єктів – ілюстрацій, питань у тестовій формі, визначень тощо, до яких у 1994 р. У. Ходжінс (Wayne Hodgins) запропонував термін «навчальний об'єкт» [7]. У дослідженнях А. Манако, О. Семеніхіної, В. Самсонової, А. Стрюка та інших розглянуто теоретичні та практичні питання використання навчальних об'єктів в освітній діяльності.

Загальні питання створення, використання, експертизи та педагогічної відповідності ЕОР дослідили у своїх працях М. Віник, А. Гуржій, І. Іванюк, О. Коневщинська, Г. Кравцов, Г. Лаврентьєва, В. Лапінський, С. Литвинова, Ю. Тарасіч, М. Шишкіна та інші.

Використанню вільного програмного забезпечення в системі освіти присвячено дослідження таких вітчизняних учених, як Є. Алєксєєв, О. Воронкін, Ю. Горошко, В. Габрусєв, Г. Злобін, М. Карпенко, М. Кияк, А. Костюченко, Л. Панченко, С. Семеріков, І. Теплицький, М. Шкардибарда та інших.

Однак, проблема розробки електронних навчальних курсів з використанням уніфікованих навчальних об'єктів, на наш погляд, потребує подальшого дослідження, у тому числі й шляхом розгляду вільного програмного забезпечення як засобу їх розробки.

**Мета статті** – дослідити засоби вільного програмного забезпечення, що використовуються для створення електронних навчальних курсів.

## 2. МЕТОДИ ДОСЛІДЖЕННЯ

Під час дослідження було використано теоретичні методи: аналіз і узагальнення наукової літератури з проблеми дослідження; системний, індуктивний, дедуктивний підходи до здійснення аналізу використання засобів вільного програмного забезпечення під час створення електронних навчальних курсів.

## 3. РЕЗУЛЬТАТИ ДОСЛІДЖЕННЯ

По-перше, електронні освітні ресурси є частиною технологічного компоненту, що разом із просторово-семантичним, інформаційним, комунікативним та імовірнісним компонентами утворюють інформаційно-освітнє середовище [3]. По-друге, без електронних освітніх ресурсів не можна створити персональне навчальне середовище, що є одним із новітніх понять електронного навчання [2]. У положенні про електронні освітні ресурси визначено, що ця категорія є складовою частиною навчально-виховного процесу, має навчально-методичне призначення, використовується для забезпечення навчальної діяльності та вважається одним із головних елементів інформаційно-освітнього середовища [4]. Виходячи з цього, маємо змогу зазначити, що створення електронних освітніх ресурсів є одним із головних завдань електронного навчання.

Поставлене завдання створення ЕОР неможливе без розгляду питання його уніфікації. Стандартизацію ЕОР розпочав у авіаційній індустрії Комітет комп'ютерного навчання в авіаційній промисловості AICC (Aviation Industry Computer based training Committee). У розробленій ним специфікації CMI Guidelines for Interoperability (CMI001) визначено взаємодію між системою дистанційного навчання і навчальним ресурсом; обмін навчальними матеріалами курсів між різними системами дистанційного навчання; збереження інформації про результати виконаних завдань. Наступний етап стандартизації полягав у створенні консорціумом IMS Global Learning Consorium (Apple, IBM, Oracle, Sun Microsystems, Microsoft, Universityof California – Berkley та інші) специфікацій [8]:
– IMS Question&Test Interoperability Specification (специфікація уніфікованих запитань і тестів);
– IMS Learning Resource Meta-data Specification (специфікація метаданих навчальних ресурсів курсу);
– IMS Content Packaging Specification (специфікація зовнішньої взаємодії обміну даними),
– IMS Learner Information Packaging (опис структури результатів навчання).

На основі стандарту IMS відповідно до ініціативи ADL (Advanced Distributed Learning) було розроблено стандарт ADL SCORM (Sharable ContentObject Reference Model) – модель обміну навчальними матеріалами. Стандарт складається з чотирьох розділів:
– Content Aggregation Mode (модель накопичення змісту);
– Run-TimeEnvironment (середовище виконання);
– Sequencing and Navigation (упорядкування та навігація);
– Conformance Requirements (вимоги відповідності).

Подальшою ініціативою ADL було створення у 2013 році стандарту Experience API (xAPI), що було розроблено під робочою назвою TinCan API. Специфікація xAPIу сфері дистанційного навчання дає змогу навчальним системам взаємодіяти між собою шляхом відстеження і запису навчальних занять усіх видів. Інформація про навчальну діяльність зберігається в спеціальній базі – сховищі навчальних записів LRS (Learning record store). LRS може бути як частиною систем дистанційного навчання, так і самостійною системою [5].

Experience API дає можливість ураховувати види навчальної активності, недоступні в SCORM, а саме: мобільне навчання, ігри, симуляції, неформальне навчання та дії тих, хто навчається за традиційними формами навчання. Experience API дає змогу отримувати дані практично з усіх існуючих електронних пристроїв:
- у яких з'єднання з Інтернетом нестабільне або є тільки в обмежений період часу;
- що використовуються в електронному навчанні;
- зі сторонніх серверів;
- з мобільних додатків, навчальних емуляторів тощо.

Середовища розробки в галузі електронного навчання орієнтовані насамперед на розробку ЕОР, що мають відповідати високим освітнім критеріям, вимагають адаптації, персоналізації і динамічної спрямованості. З безперервним збільшенням джерел інформації і розроблених навчальних ресурсів стає все важче контролювати й актуалізувати ЕОР, рекомендувати ті, що найбільш відповідають конкретній особі та конкретним освітнім цілям. Саме тому розробка в ручному режимі навчального змісту і безпосереднє управління процесом навчання все частіше поступаються місцем розробці програмного забезпечення для автоматизації створення ЕОР, співпраці в процесі розробки й автоматичної їх адаптації в конкретному контексті. Отже, однією з основних тенденцій розвитку середовищ електронного навчання є вдосконалення інструментів їх програмної розробки і, врешті-решт, їх перетворення на інтегровані середовища для розробки й управління навчальними ресурсами.

Нами було практично досліджено стан проблеми використання існуючого вільного програмного забезпечення в процесі створення електронних навчальних курсів у рамках навчальної дисципліни «Сучасні інформаційні технології в навчанні», що викладається студентам магістратури фізико-математичного факультету ДВНЗ «Донбаський державний педагогічний університет». До переліку вільного програмного забезпечення, що було запропоновано для вивчення, віднесено Xerte Online Editor, eXee Learning XHTML editor та Reload Editor.

Xerte Online Toolkits (XOT) – програмне забезпечення, що містить інструменти для розробки засобів електронного навчання розроблене в академічному середовищі університету Ноттінгем, розповсюджене за ліцензією Apache License та. Головне призначення повнофункціонального середовища полягає у створенні інтерактивних об'єктів навчання. Існуюча на сьогодні версія 3.4 дає можливість створення сучасних навчальних курсів із доволі складною структурою і різноманітними, у тому числі й інтерактивними, навчальними об'єктами, і не потребує поглиблених знань в галузі програмування. Для розробки такого курсу автору достатньо використовувати лише браузер, а всі операції зі створення електронних освітніх курсів виконуються за допомогою інтуїтивно зрозумілих дій.

У системі XOT закладено можливість створювати електронні ресурси таких чотирьох видів: інтерактивний навчальний курс; web-сторінка з мультимедійними елементами; ресурс для тестування; канал новин. Кожен з означених видів має широкі можливості. Так, наприклад, для інтерактивного навчального курсу існує можливість додавати текстові, медійні, навігаційні, інтерактивні елементи, а також елементи ділової графіки, ігри, посилання на відеохостинги тощо (рис. 1). Кожен із наявних розділів має достатню кількість різноманітних елементів, що відповідають як сучасному рівню розвитку web-технологій (зокрема HTML5), так і сучасному уявленню про необхідність і можливість їх використання у процесі створення електронних освітніх ресурсів. Попри це, кожен елемент має досить представницьку кількість параметрів і опцій. Вбудований текстовий редактор дає можливість не тільки задавати атрибути тексту, але й вбудовувати велику кількість наявних графічних примітивів інформаційного змісту, медіа-контент різноманітних доступних сайтів і їх частини, flash-анімацію; розбивати інформацію на блоки та перевіряти орфографію введеного тексту.

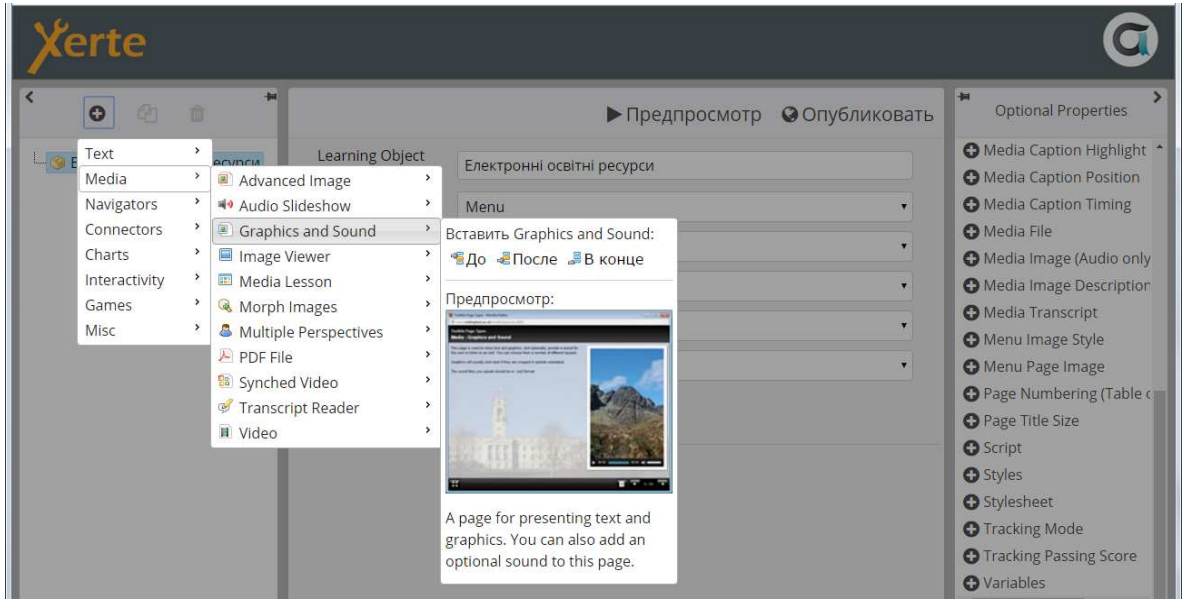

*Рис. 1. Створення інтерактивного навчального курсу в системі XOT*

У процесі створення web-сторінки з мультимедійними елементами використовується значно менша кількість готових елементів, однак створена сторінка відповідає всім необхідним параметрам для перегляду її на будь-якому пристрої, включаючи мобільні, і має гнучку систему налаштування зовнішнього вигляду ресурсу (рис. 2.).

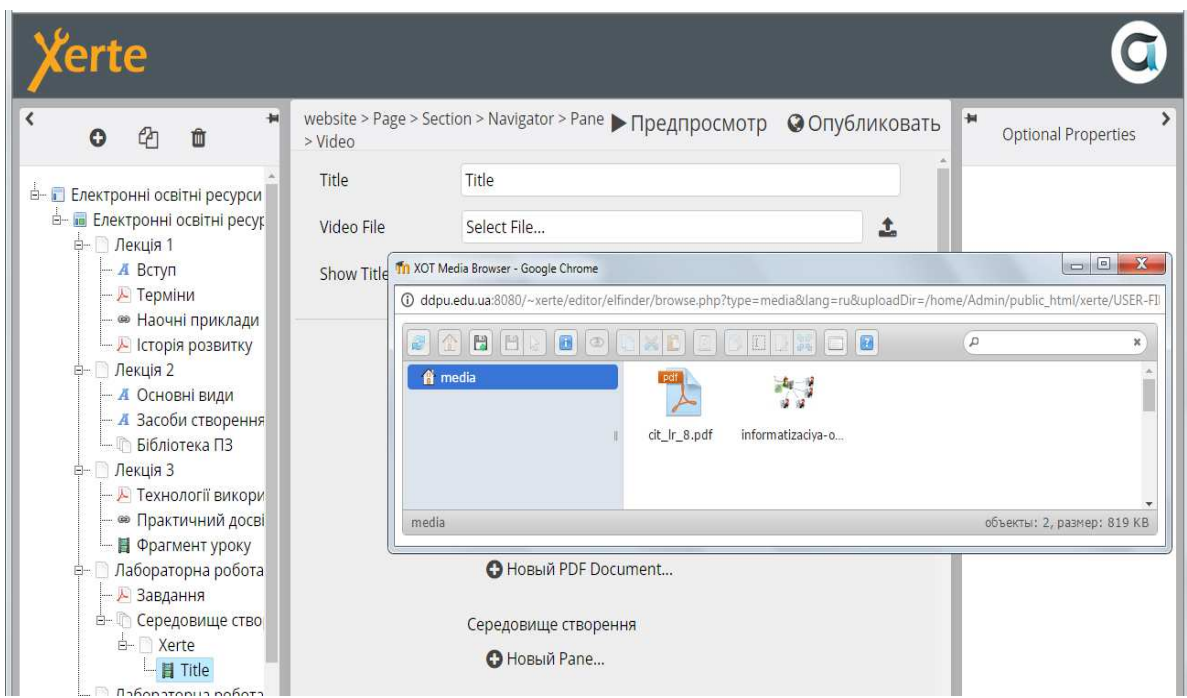

*Рис. 2. Розробка web-сторінки в системі XOT*

У створенні ресурсу для тестування використовуються такі види тестів: множинний вибір, вибір числового значення, вільна відповідь і закрита відповідь. Подібно до попередніх двох видів існує можливість налаштування як самого ресурсу для тестування, так і опцій тестових завдань. Останнім видом ресурсів, розроблених системою XOT, є створення каналу новин RSS.

Використовуючи сучасні технології, система Xerte Online Toolkits дає можливість вбудовувати в навчальний курс такі елементи, як відео-, фото- та аудіоресурси, діаграми, графіки, схеми, flash-анімацію та інтерактивні елементи керування, має сумісність із системою LaTeX для створення математичних текстів.

Створені в системі XOT курси можна експортувати до інших систем електронного навчання, використовуючи такі стандарти, як SCORM 1.2, SCORM 2004 та IMS Content Package.

eXe (eLearning XHTML editor) є web-інструментом розробки, призначеним для надання допомоги викладачам у галузі проектування, розробки та публікації web-орієнтованих навчально-методичних матеріалів. Система eXe також розробляється в академічному середовищі Університету Окленда та Оклендського університету технологій. Вона може генерувати інтерактивний навчальний матеріал у форматі XHTML або HTML5 і дає можливість створювати навчальні ресурси, що містять текст, зображення, інтерактивні компоненти, галереї зображень або мультимедійні кліпи. Такі файли можуть бути експортовані в різних цифрових форматах, що будуть використовуватися незалежно один від одного на веб-сайті інструктора, або інтегруватися в систему управління навчанням (LMS). Поточна версія 2.1 підтримує такі формати, як IMS Content Package, SCORM 1.2, SCORM 2004, IMS Common Cartridge formats, ePub3 або web-ресурс у форматі HTML5.

Навчальний курс визначається структурою, що в системі eXe відображено у вигляді дерева (за замовчуванням Головна>Тема>Секція>Одиниця) та інтерактивних елементів (iDevices), які користувач може розмістити на будь-якому рівні курсу. Кожен елемент курсу є сукупністю структурних одиниць, що описують зміст навчання та є готовими шаблонами змісту навчання, до яких віднесено, наприклад, такі компоненти, як Галерея зображень, Мульти-вибір, Зовнішній Web ресурс, JavaАплет, SCORM Quiz, Wiki-Стаття та інші структурні одиниці (рис. 3). Кожен інтерактивний елемент має поради щодо його використання, що сприяє визначенню доцільності його введення до навчального курсу. Попри йе, кожен із параметрів інтерактивного елементу має відповідні пояснення його значення і способи використання. Розглянемо, наприклад, елемент «Вправа «Пропущені Слова» (рис. 4-5). Як змінні елементи доступні: заголовок інтерактивного елементу, інструкційний текст до завдання, текст самого завдання (текст, що необхідно ввести, виокремлюється за допомогою кнопки «Сховати/Показати Слово»), текст зворотного зв'язку, що виводиться після введення відповіді.

Під час електронного навчання необхідним є використання компонентів діагностики отриманих знань. Система eXe пропонує використовувати для такої діяльності елементи «Питання типу Правильно/неправильно», «SCORM Quiz», «FPD – Close Activity», «Мульти-вибір», «Мульти-можливість» тощо. Наведені компоненти дають можливість створювати різноманітні за своїм призначенням інтерактивні елементи курсу.

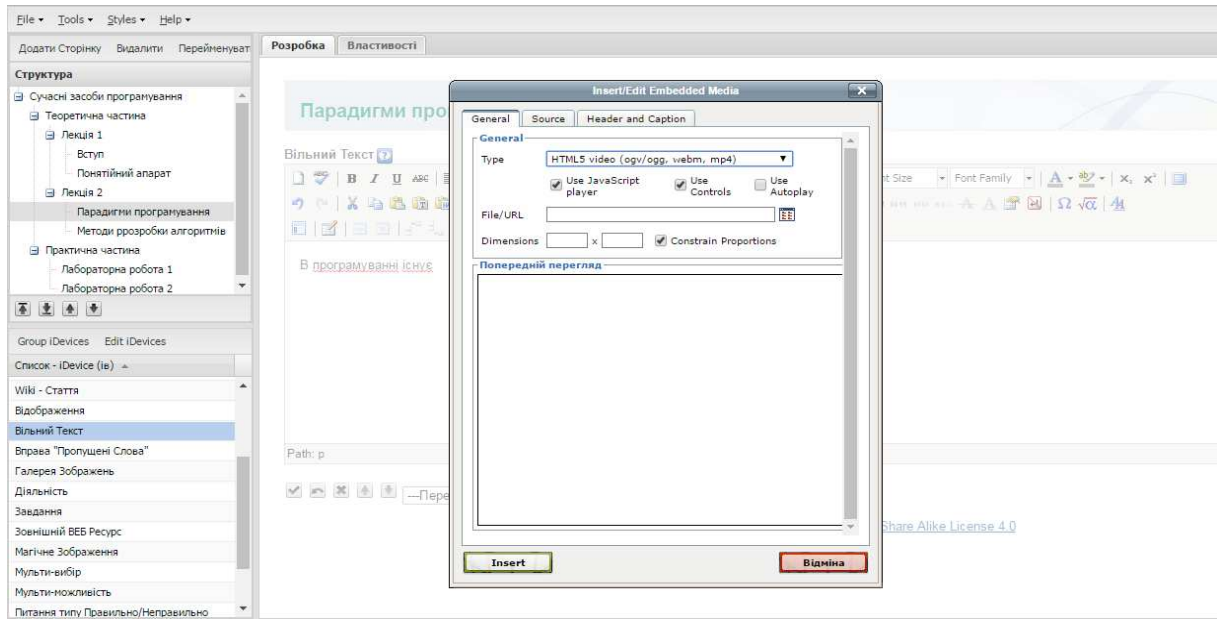

*Рис. 3. Створення навчального курсу в системі eXe*

Велика кількість інтерактивних елементів системи eXe передбачає використання можливостей стандартних текстових редакторів. Попри це, ця система має власний потужний редактор, що за своїми функціональними можливостями співвідноситься з RichTextEditor (рис. 6).

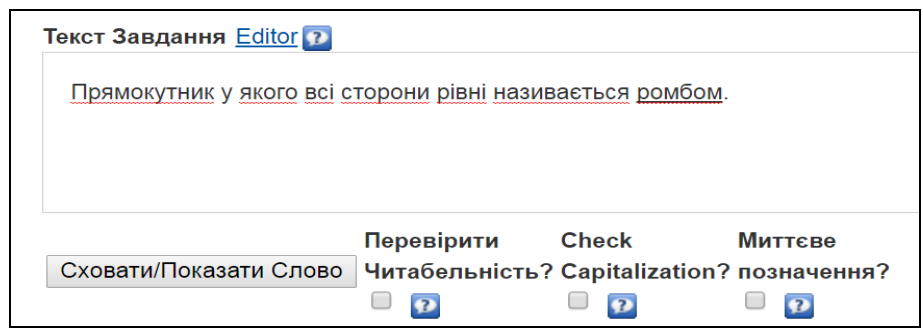

*Рис. 4. Режим створення інтерактивного елемента*

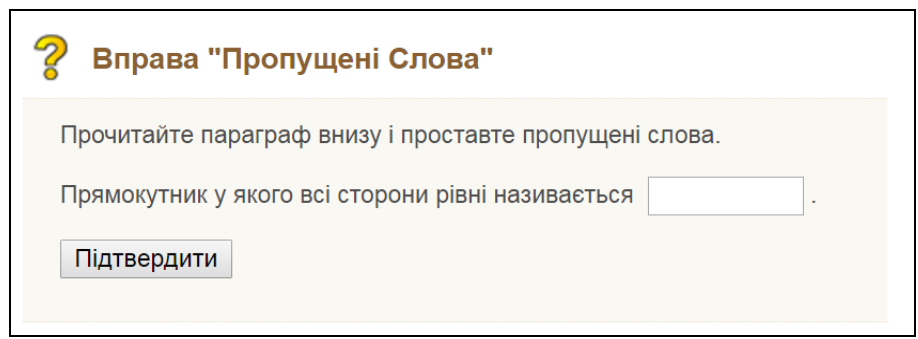

*Рис. 5. Режим використання інтерактивного елементу*

Наступні типи медіа-ресурсів можуть бути розміщені в навчальному курсі через використання вбудованого редактора: HTML5 video (ogv/ogg, webm, mp4, wave), Flash (swf, flv, mp3), QuickTime (mov, mpeg), Windows Media (wmv, wma), RealMedia (ram) та iframe, що можуть бути створеними за допомогою CamStudio, LiVES, Shotcut, Kdenlive, OpenShot, Pitivi. За умови розміщення зазначених ресурсів є можливість налаштування власних атрибутів, доступних у вкладці «Додатково». Попри це, редактор дає можливість вбудовувати сучасні конструкції структурованого розміщення інформації, таких як Accordion, Tabs, Pagination, Carousel та Timeline. Використовуючи запропоновані конструкції, розробник курсу може у компактному вигляді в рамках одного екрану розміщувати навчальний матеріал, поділений на підрозділи. Додавання математичних виразів виконується через створення графічного зображення з вхідного коду, що відповідає командам LaTeX. Існує можливість вибору візуалізованих математичних символів і літер грецького алфавіту у вікні створення математичних виразів.

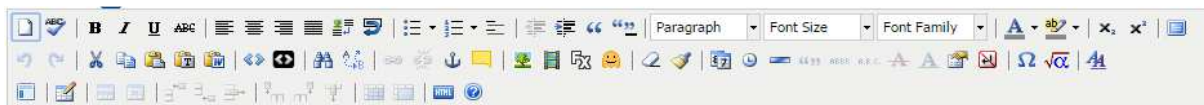

*Рис. 6. Вбудований редактор у систему eXe*

Окремо варто зробити зауваження про такий компонент, як «JAVA Аплет». Окрім варіанту власноручного створення аплету або завантаження його з файлу, існує можливість додати аплети, створені в таких системах навчання, як Descartes, Geogebra, JClic та Scratch.

Позитивним моментом системи eXe є її локалізація. Хоча вона виконана не повністю і з помилками, однак, дає змогу оформити навчальні курси мовою, якою створено навчальний курс.

У разі, коли навчальні об'єкти вже створено, є можливість скористатись таким програмним засобом, як ReloadEditor. Він створюється під егідою Joint in formation services committee та розробляється в університетах Болтона та Стратклайда в рамках програми Exchange for Learning Programme (X4L). Reload Editor дає можливість створювати розвинуті ієрархічні курси з можливостями редагування метаданих користувачем (рис. 7). Редактор має вбудований емулятор LMS, за рахунок чого можна переглядати результати створеного курсу в дії згідно зі стандартами SCORM та IMS. Програмний засіб Reload Editor розроблений мовою програмування Java, що визначає як переваги, так і недоліки його використання.

Навчальний курс, як і в попередніх програмних засобах, подано у вигляді ієрархічного дерева з навчальними об'єктами й може бути відредаговано на будь-якому етапі розробки. До негативних моментів системи слід віднести відсутність вбудованих механізмів створення навчальних об'єктів і відсутність локалізації.

Навчальна дисципліна «Сучасні інформаційні технології в навчанні», що викладається на фізико-математичному факультеті ДВНЗ «ДДПУ», має на меті надання майбутнім фахівцям теоретично обґрунтованих знань і наочно сформованих умінь використання сучасних інформаційних технологій в навчальній діяльності; підготовку до самоосвітньої діяльності та самовдосконалення, а тому лекційний матеріал містить, окрім операційних відомостей про системи, що розглядаються, також і методичну складову використання результатів їх роботи. Так, для створених магістрантами електронних навчальних курсів відповідно до навчальної програми дисципліни було проаналізовано їх дидактичні можливості. Результати аналізу засвідчили, що обране вільне програмне забезпечення дало змогу створити курси, що урізноманітнюють форми подання навчальної інформації; урізноманітнюють типи навчальних завдань; забезпечують реакцію на дії тих, хто навчається; індивідуалізують процес навчання, використання основних і допоміжних навчальних впливів; застосовують ігрові прийоми; відтворюють фрагменти навчальної діяльності; активізують навчальну роботу, посилюють мотивацію до навчання. Магістрантами були створені електронні навчальні курси: «Ланцюгові дроби», «Жорданова нормальна форма», «Крайові задачі», «Системи комп'ютерної математики», «Дистанційні системи навчання», «Сучасні інформаційні технології», «Основи криптографії», «Культурологія», «Візуальні технології програмування», «Комбінаторика», «Задачі на побудову», «Афінна геометрія».

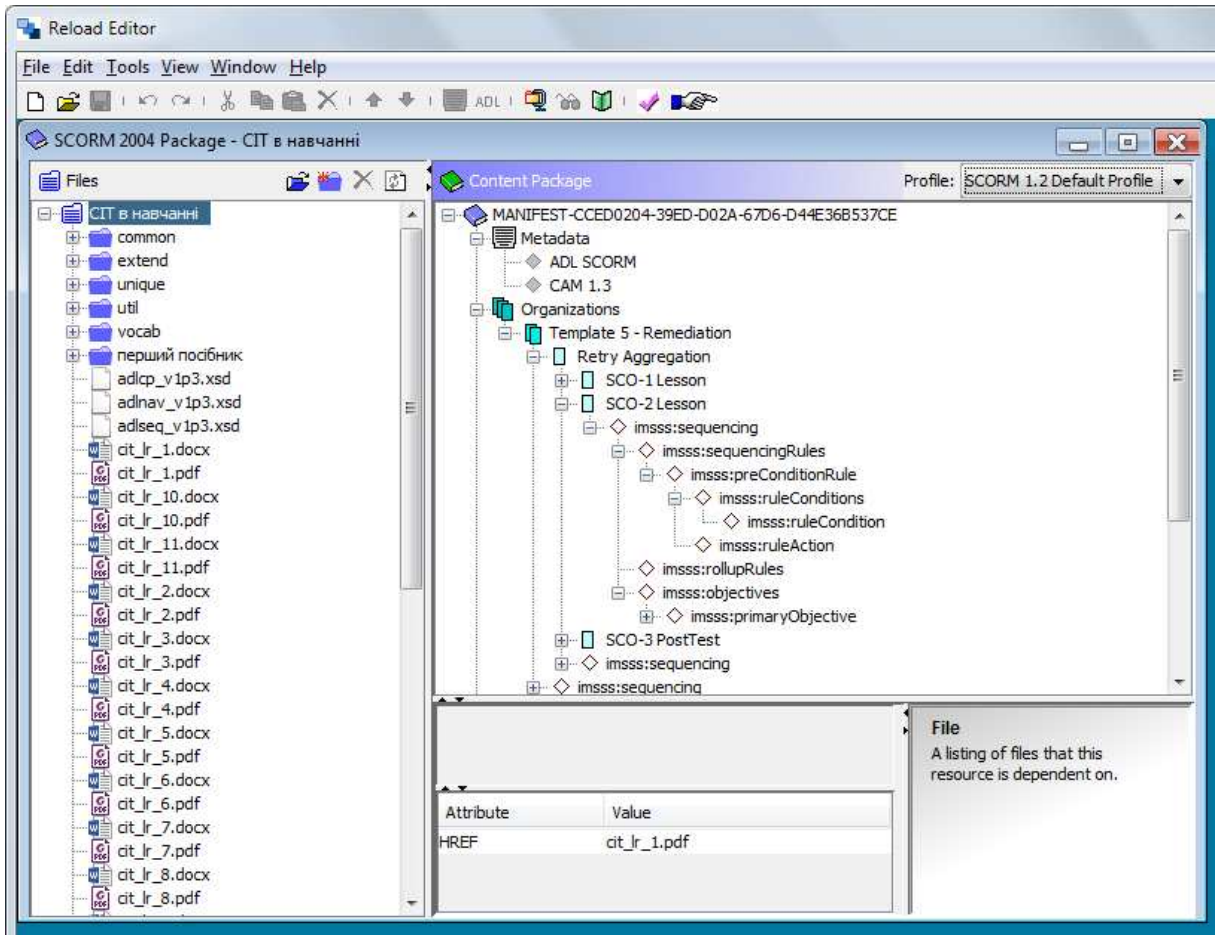

*Рис. 7. Розробка навчального курсу в системі ReloadEditor*

Упровадження створених електронних навчальних курсів відбувалося двома шляхами. Для навчальних дисциплін, що використовують змішану форму навчання, курси було імпортовано до системи дистанційного навчання фізико-математичного факультету. Ті ж навчальні дисципліни, що викладаються за традиційною формою, використовували електронні навчальні курси як навчальний матеріал під час самостійної роботи.

Результати впровадження надали можливість отримати більш детальне уявлення про можливості систем ХОТ, eXe та Reload Editor. Зазначені системи розглянуто нами з точки зору їх взаємодії як із системами дистанційного навчання, так і з позиції доступності створених ресурсів. Так, кожна із систем відповідає стандартам SCORM та IMS обміну навчальними об'єктами. Важливим фактором у виборі системи для розробки електронних навчальних курсів є можливість і наявність її локалізації. Найкращий варіант локалізації із запропонованих має система eXe, а ReloadEditor, на даний час, взагалі не локалізований. Попри це, нами враховано повноцінну функціональність роботи усіх систем у процесі створення електронних навчальних курсів, і за цим показником система ReloadEditor не є самостійною. Оскільки кожна з систем має різні платформи розробки та функціонування, то й порівнювати їх важко. Перевагою ХОТ є її доступність як сервісу мережі, перевагою ж системи eXe є відсутність необхідних додаткових компонентів платформи. Попри це, перевагою ХОТ слугує наявність відкритого навчального контенту на сайті розробника програмного продукту. Різниться також і кількість убудованих навчальних об'єктів для розробки електронних навчальних курсів. Безперечну перевагу саме за цим фактором має система ХОТ, але при цьому обмежені можливості імпорту ззовні. Системи eXeі Reload Editor на відміну від ХОТ мають різноманітні можливості щодо імпорту навчальних об'єктів ззовні. Узагальнені результати дослідження засобів вільного програмного забезпечення для створення електронних освітніх курсів представлено нами в таблиці 1.

Як уже було зазначено, кожна із систем створення електронних навчальних курсів розроблена на різних платформах, а тому, і їх використання дещо відрізняється. СистемуХОТ створено як мережевий сервіс і відповідно дана система надає доступ до розроблених електронних навчальних курсів як до мережевого ресурсу. Система eXe працює як окремий додаток, однак через мережеві протоколи останнє надає можливість створення мережевих електронних навчальних курсів, але для їх публікації необхідно використовувати додаткові засоби. Те саме стосується й системи ReloadEditor, до переваги якої можна віднести кросплатформеність, а до недоліків, на що, під час виконання лабораторних робіт, скаржилися і студенти, – відсутність інтуїтивно зрозумілої операційної діяльності (відсутність інтуїтивно зрозумілих послідовних дій).

*Таблиця 1.*

**Порівняння систем створення електронних навчальних курсів**

| Параметри | Xerte Online Toolkits (XOT) | eXe (eLearning XHTML editor) | Reload Editor |
|---|---|---|---|
| Стандарти обміну | SCORM 1.2, SCORM 2004, IMS Content Package | IMS Content Package, SCORM 1.2, SCORM 2004, IMS Common Cartridge formats | IMS Content Package, SCORM 1.2, SCORM 2004 |
| Необхідність додаткової платформи | XAMPP | Ні | JRE |
| Можливість автономної розробки навчальних об'єктів | Так | Так | Ні |
| Кількість навчальних об'єктів | > 60 | > 25 | - |
| Можливість імпорту навчальних об'єктів | Частково | Так | Так |
| Локалізація | Частково | Так | Ні |
| Останнє оновлення | Лютий 2017 | Березень 2017 | Жовтень 2013 |
| Відкритий навчальний контент | Так | Ні | Ні |

## 4. ВИСНОВКИ ТА ПЕРСПЕКТИВИ ПОДАЛЬШИХ ДОСЛІДЖЕНЬ

Виконане дослідження засобів вільного програмного забезпечення, що використовується для створення електронних навчальних курсів, дає нам змогу зробити висновки про те, що такі програмні продукти мають достатній функціонал для їх використання. Додатковим підтвердженням висновку є підтримка розглядуваним програмним забезпеченням стандартів обміну навчальними об'єктами, що свідчить про перспективи використання вільного програмного забезпечення в електронній освіті.

У результаті дослідження було виявлено, що система Xerte Online Toolkits є самодостатнім засобом вільного програмного забезпечення, за допомогою якого створюються електронні навчальні курси. Арсенал системи має значну кількість шаблонів навчальних об'єктів, а створений курс може бути опубліковано в мережі безпосередньо із системи ХОТ. За допомогою системи eLearning XHTML editor, що має значно меншу кількість шаблонів навчальних об'єктів, можливо створювати електронні навчальні курси різного стильового оформлення. Для цього не потрібно додаткових засобів розробки до того ж система eXe доцільна для використання в локальному варіанті. Остання розглянута нами система ReloadEditor призначена для напівавтоматичного створення електронних навчальних курсів, які складаються з розроблених ззовні навчальних об'єктів. Дана система доцільна у використанні професійними командами розробників.

Результати практичного використання засобів вільного програмного забезпечення в процесі створення електронних навчальних курсів показали недостатність розробленості методики створення електронних начальних курсів, особливо з урахуванням нових можливостей інформаційно-комунікаційних технологій, та їх використання як засобу електронного навчання.

# СОЗДАНИЕ ЭЛЕКТРОННЫХ ОБРАЗОВАТЕЛЬНЫХ КУРСОВ СРЕДСТВАМИ СВОБОДНОГО ПРОГРАММНОГО ОБЕСПЕЧЕНИЯ

**Величко Владислав Евгеньевич**
канд. физ-мат наук, доцент
ГВУЗ «Донбасский государственный педагогический университет», г. Славянск, Украина
ORCID ID 0000-0001-9752-0907
*vladislav.velichko@gmail.com*

**Аннотация.** Информатизация образования обогащает традиционные методики обучения новыми формами и методами, которые основаны на широком и гармоничном использовании ИКТ. Электронная педагогика приобретает статус одного из самых востребованных направлений развития педагогики нашего времени, в котором такое средство как электронные учебные курсы является результатом аккумулирования современных информационно-коммуникационных технологий с теорией и практикой электронного обучения. В статье рассматриваются стандарты обмена информацией между системами обучения и свободное программное обеспечение, используемое для создания электронных учебных курсов. Отмечено, что электронные учебные курсы являются как частью дистанционных систем обучения, так и самостоятельным средством. Рассмотренные системы создания электронных учебных курсов позволяют самостоятельно или частично создавать электронные образовательные ресурсы, опираясь на современное видение объектной модели представления информации.

**Ключевые слова:** электронное обучение; свободное программное обеспечение; электронные учебные курсы; Xerte Online Toolkits; eLearning XHTML editor; Reloa dEditor.

# CREATING E-LEARNING MEANS OF FREE SOFTWARE

**Vladyslav Ye. Velychko**
PhD (Physical and Mathematical Sciences), associate professor
Donbas State Pedagogical University, Sloviansk, Ukraine
ORCID ID 0000-0001-9752-0907
*vladislav.velichko@gmail.com*

**Abstract.** Informatization of education enriches traditional teaching methods with new forms and methods, which are based on the broad and harmonious use of ICT. Electronic pedagogy acquires the status of one of the most popular trends in the development of pedagogy of our time, in which such a tool as e-learning courses is the result of the accumulation of modern information and communication technologies with the theory and practice of e-learning. The article examines the standards of information exchange between learning systems and free software used to create e-learning courses. It is noted that electronic training courses are part of distance learning systems, as well as an independent tool. The considered systems of creation of electronic training courses allow creating electronic educational resources independently or partially, relying on the modern vision of the object model of information representation.

**Keywords:** e-learning; free software; e-learning courses; Xerte Online Toolkits; eLearning XHTML editor; Reload Editor.

## REFERENCES (TRANSLATED AND TRANSLITERATED)